\documentclass[final,1p,times]{elsarticle} 

\usepackage{graphicx}
\usepackage{amssymb} 
\usepackage{amsthm} 
\usepackage{lineno}
\usepackage{subfigure}

\newcommand{\kt}{{$k_{\rm T}$}}

\newcommand{\PbPb}{Pb--Pb }

\newcommand{\sNN}{\ensuremath{\sqrt{s_\mathrm{NN}}}}
\newcommand{\GeV}{GeV}
\newcommand{\GeVc}{\GeV/\ensuremath{c}}

\newcommand{\rom}[1]{{\mathrm{#1}}}   
\newcommand{\pt}{\ensuremath{p_\rom{T}}}
\newcommand{\deltapt}{\ensuremath{\rom{\delta}\pt}}

\newcommand{\ptconst}{\ensuremath{\pt^\rom{const}}}

\newcommand{\Fig}[4]{
    \begin{figure}[!h]
        \centering
        \includegraphics[#3]{#1}
        \caption{\label{#2}\small#4}
    \end{figure}
}

\newlength{\myfigwidth}
\newlength{\mydoublefigwidth}
\newlength{\figurematrixwidth}
\journal{Nuclear Physics A} 

\begin{document}

\begin{frontmatter} 

\title{Measurement of jet spectra with charged particles in Pb-Pb collisions at
$\sqrt{s_{NN}}$=2.76 TeV with the ALICE detector}

\author{Marta Verweij\fnref{fn1} (for the ALICE\fnref{col1} Collaboration)}
\fntext[col1] {A list of members of the ALICE Collaboration and acknowledgements can be found at the end of this issue.}
\address[fn1] {Utrecht University}
\ead{marta.verweij@cern.ch}


\begin{abstract} 
  We report a measurement of transverse momentum spectra of jets detected with
  the ALICE detector in Pb-Pb collisions at $\sNN=2.76$ TeV. Jets are
  reconstructed from charged particles using the anti-\kt\ jet algorithm. The
  transverse momentum of tracks is measured down to $150$ MeV/$c$ which gives
  access to the low \pt\ fragments of the jet. The background from soft
  particle production is determined for each event and subtracted. The
  remaining influence of underlying event fluctuations is quantified by
  embedding different probes into heavy-ion data. The reconstructed transverse
  momentum spectrum is corrected for background fluctuations by unfolding. We
  observe a strong suppression in central events of inclusive jets
  reconstructed with radii of $0.2$ and $0.3$. The fragmentation bias on jets
  introduced by requiring a high \pt\ leading particle which rejects jets with
  a soft fragmentation pattern is equivalent for central and peripheral
  events.
\end{abstract} 


\end{frontmatter} 


Jets are a powerful tool to study the properties of the medium created in
heavy-ion collisions. The kinematic properties of the jets reflect the
kinematic properties of the original hard partons from the hard process. It is
expected that the kinematic properties of jets are modified in the presence of
a medium. The challenge in heavy-ion collisions is to disentangle
the jet fragments originating from hard scatterings from the very large
background due to soft processes.

This analysis uses data collected by the ALICE experiment in the heavy-ion run
of the LHC in the fall of $2010$ with an energy $\sNN = 2.76$ TeV.
Jets are clustered from charged particles measured in the central
tracking detectors: Inner Tracking System (ITS) and Time Projection Chamber
(TPC). This ensures a uniform acceptance in full azimuth and $|\eta|<0.9$ with
high tracking efficiency.

For the clustering of jet candidates the anti-\kt\ algorithm
\cite{Cacciari2008b} with resolution parameters $R=0.2$ and $R=0.3$ is
used. We subtract from each jet candidate event-by-event the average
background. The minimum \pt\ of the jet constituents is $0.15$ GeV/$c$. All
jets with a jet axis within $|\eta|<0.5$ are considered for this analysis.
The \kt\ algorithm \cite{Cacciari2006} is used to obtain a collection of
background clusters in each event from which the average transverse momentum
per unit area $\rho$ is calculated by taking the median of $(p_{\rom{T,j}}/A_{\rom{j}})$
(where $A_{j}$ is the area of the background cluster) of all \kt\ clusters. To
reduce the contribution from true hard jets the two leading \kt\ clusters are
excluded from the calculation of $\rho$. The estimated background momentum
$\rho \cdot A$ is subtracted from the reconstructed \pt\ of each anti-\kt\ jet
candidate in the event \cite{Cacciari:2007fd, Cacciari2010}.

Point-to-point fluctuations of the background are quantified by placing random
cones in the measured \PbPb events and by embedding high
\pt\ probes \cite{Abelev:2012ej}. The reconstructed transverse momentum
$\pt^{\rom{rec}}$ of the embedded probe in the heavy-ion environment is
compared to the embedded transverse momentum $\pt^{\rom{probe}}$ by
calculating the difference: $\deltapt = \pt^{\rom{rec}} - \rho \cdot
A_{\rom{jet}} - \pt^{\rom{probe}}$. Fluctuations of the background depend
strongly on the multiplicity, jet area (or radius) and minimum \pt\ of the jet
constituents \ptconst. Background fluctuations have a large impact on the
measured jet spectrum due to the finite probability for large positive
flucuations. The width of the background fluctuations, $\sigma(\deltapt)$, for
$10$\% most central events and $\ptconst>0.15$ \GeVc\ is $4.47$ \GeVc\ for
$R=0.2$ jets and $7.15$ \GeVc\ for $R=0.3$ jets. Background fluctuations are
corrected for statistically via unfolding. Combinatorial jets consisting of a
random collection of particles which do not originate from a hard process are
removed in the unfolding by not constraining the region below
$\pt^{\rom{measured}}=30$ \GeVc\ with measured data. In addition, jet spectra
are also extracted by requiring a minimum \pt\ of the leading track in the
jet. The requirement of a high \pt\ leading track removes a large part of the
combinatorial jets in the sample while introducing a bias to harder
fragmentation. Jets with a soft fragmentation pattern are removed from the
sample when a high \pt\ leading track is required.

\section*{Results}
\begin{figure}[hp]
  \centering
  \subfigure[Centrality $0-10$\%]{\label{fig:UnfSpectraHadTrigCent0_R02}
    \includegraphics[width=0.45\linewidth]{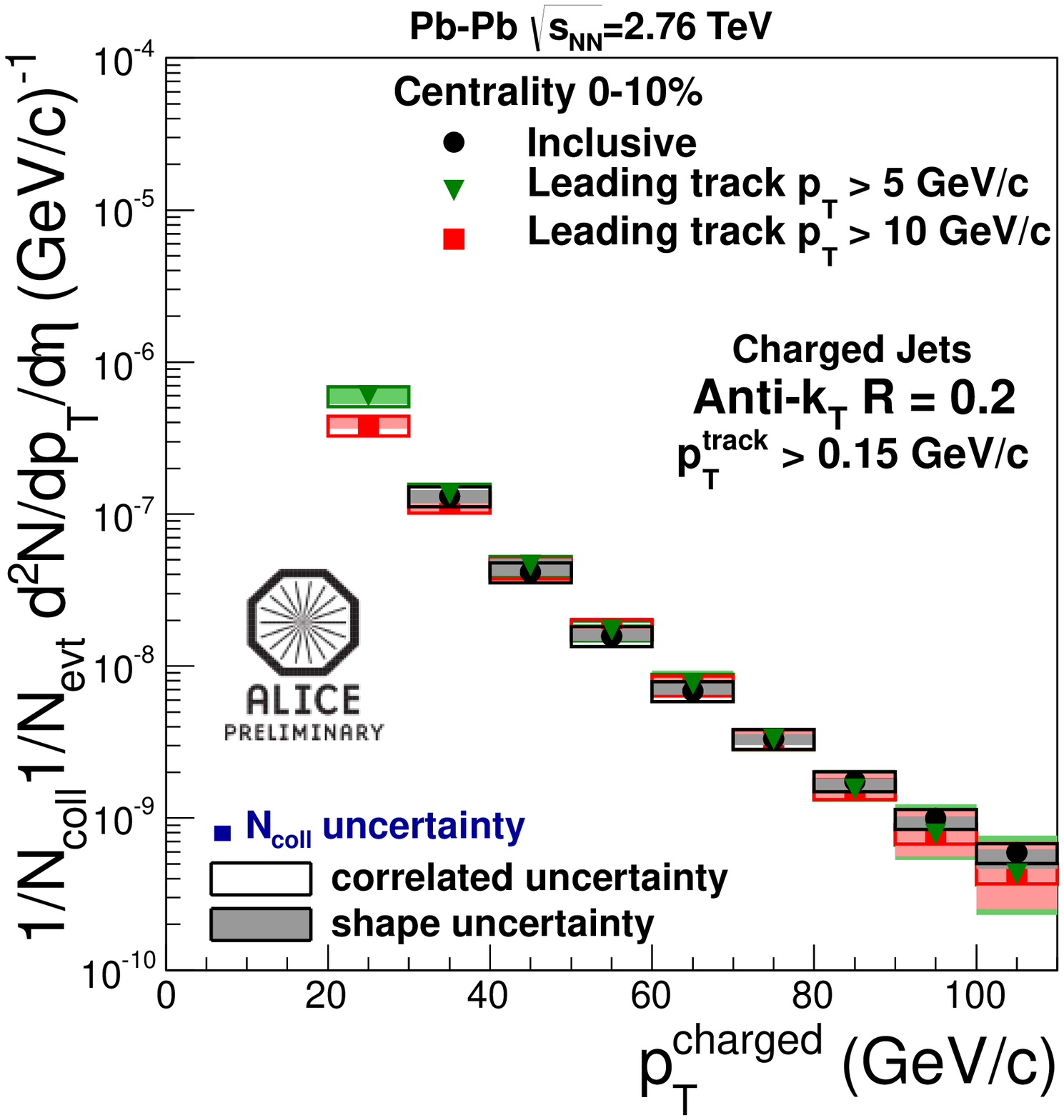}}
  \hspace{.1in}
  \subfigure[Centrality $50-80$\%]{\label{fig:UnfSpectraHadTrigCent3_R02}
    \includegraphics[width=0.45\linewidth]{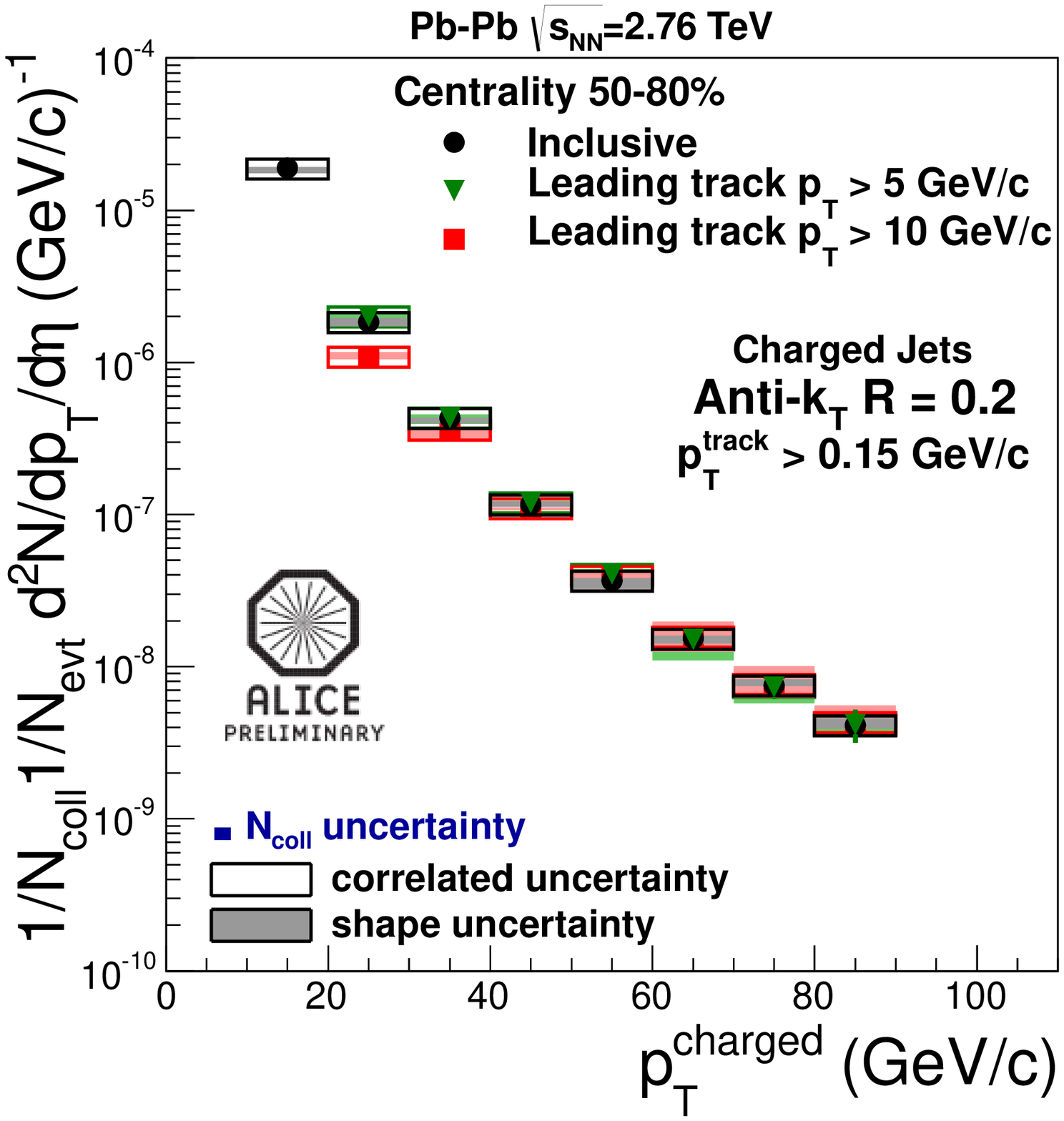}}
  \caption{Inclusive jet spectra with no
  requirement on the leading track and jets with a leading track of at least
  $5$ and $10$ GeV/$c$.\label{fig:UnfSpectraR02HadTrig}}
\end{figure}
Jet spectra are unfolded using a $\chi^{2}$ minimization method which
minimizes the difference between the unfolded spectrum convoluted with the
response matrix (the refolded spectrum) and the measured spectrum. 
The $\chi^{2}$ function used in this analysis is:
\begin{equation}\label{eq:chi2function}
\chi^{2} =
\displaystyle\sum_{\mathrm{refolded}}\left(\frac{y_{\mathrm{refolded}}-y_{\mathrm{measured}}}{\sigma_{\mathrm{measured}}}\right)^{2}
+ \beta\displaystyle\sum_{\mathrm{unfolded}}\left(\frac{\mathrm{d}^{2}\log y_{\mathrm{unfolded}}}{\mathrm{d}
  \log \pt^{2}}\right) ^{2},
\end{equation}
in which $y$ is the yield of the refolded, measured, or unfolded jet spectrum
and $\sigma_{\rom{measured}}$ the statistical uncertainty on the measured jet
spectrum.
The first summation term of equation \ref{eq:chi2function} gives the
$\chi^{2}$ between the refolded spectrum and the measured jet spectrum and the
second summation term regularizes the unfolded solution favoring a local power
law.

The response matrix includes the smearing of the measured background
fluctuations and detector effects which are determined from event and detector
simulations \cite{ALICE:2012ch}. When a leading track in the jets is
required, a small bias due to collective flow is introduced: the average
\pt\ density $\rho$ of the full \PbPb\ events is at maximum $3$ \GeVc\ smaller
than the local \pt\ density in the region of the event where a jet with a high
\pt\ track is present. For the cross-section of jets with radius $R=0.2$
this effect is negligible and no extra correction is required.

The corrected jet spectra with jet radius $R=0.2$ are presented in Figure
\ref{fig:UnfSpectraR02HadTrig} for central (0-10\%) and peripheral collisions
(50-80\%) \cite{RReedQM2012}. For both centrality classes the unbiased and leading track
\pt\ biased spectra are shown. Requiring a leading high \pt\ track in the jet
does not modify the jet sample at high \pt\ while, as expected, at low \pt\ the
jet yield is reduced.

\Fig{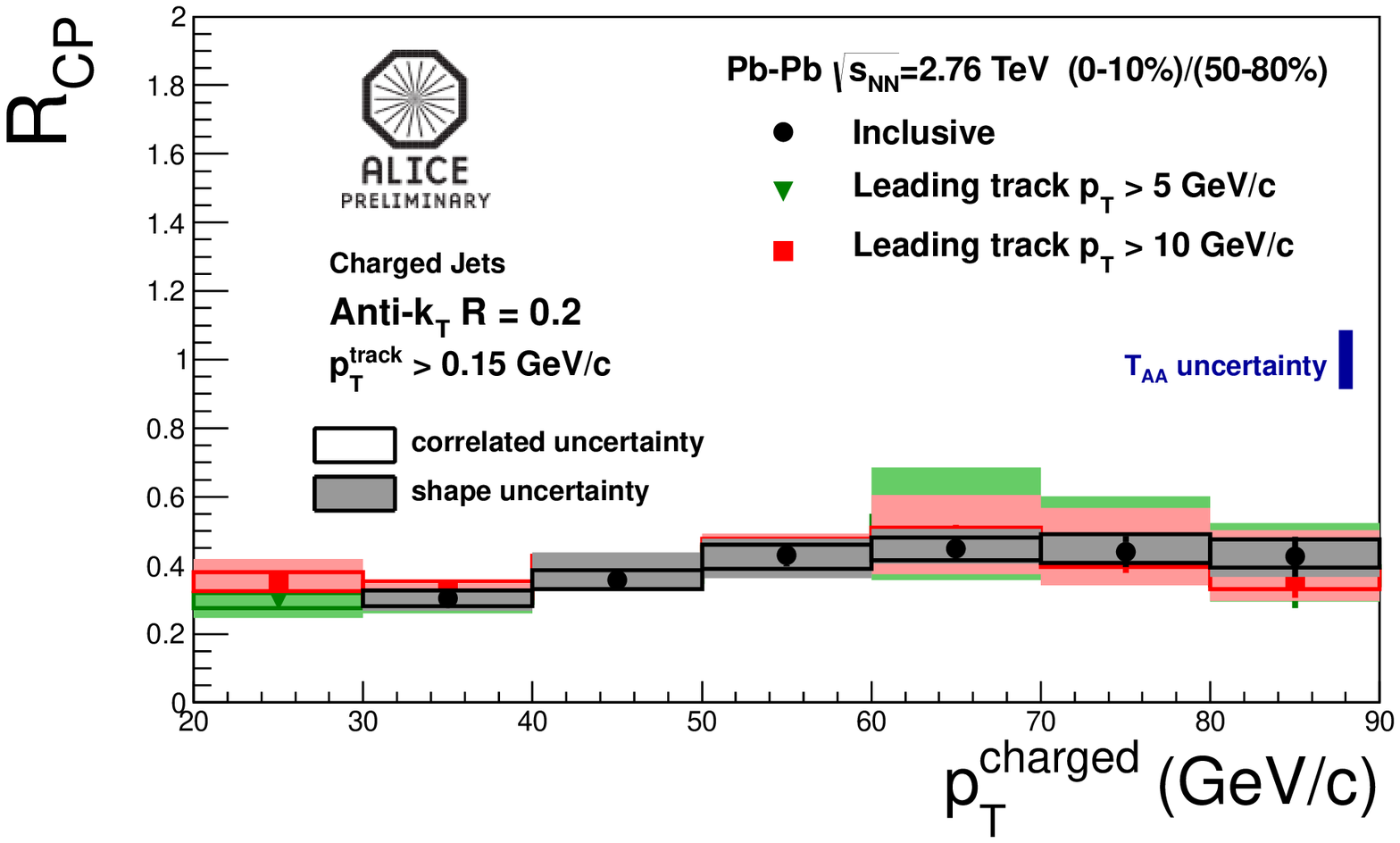}{fig:RCPHadTrig}{width=0.69\linewidth}{
  Jet nuclear modification factor $R_{\rom{CP}}$ for inclusive jet spectra with no
  requirement on the leading track and jets with a leading track of at least
  $5$ and $10$ GeV/$c$.
}
Figure \ref{fig:RCPHadTrig} shows the nuclear modification factor of jets,
$R_{\rom{CP}}$, in central collisions with respect to peripheral collisions. A
strong suppression which does not depend on the leading track requirement is
observed. The fragmentation bias due to the leading track requirement of
$\pt>5$ and $\pt>10$ \GeVc\ is observed to be similar in central and
peripheral collisions for $p_{\rom{T,jet}}^{\rom{ch}}>30$ \GeVc. The strong jet
suppression, $R_{\rom{CP}} \simeq 0.4$, implies that the full jet energy is not
captured by jet reconstruction in heavy-ion events. This is consistent with
out-of-cone radiation induced by the interaction of the parton with the dense
medium.

\Fig{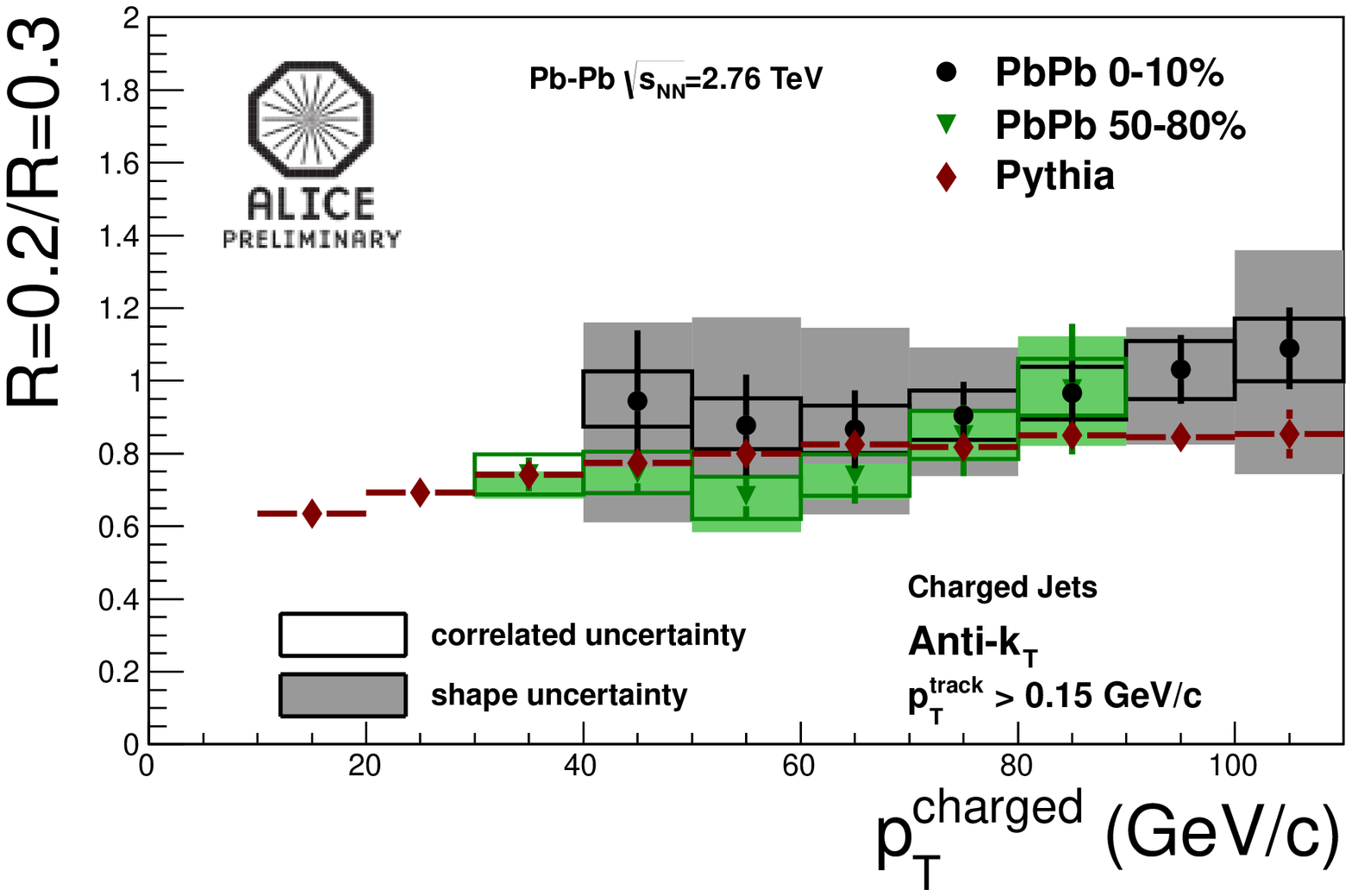}{fig:RatioR02R03CompPythia}{width=0.68\linewidth}{
  Ratio between meastured cross-sections with radii $R=0.2$ and $R=0.3$ in
  \PbPb\ collisions for central and peripheral events compared to PYTHIA \cite{Sjostrand2006}.
}
Figure \ref{fig:RatioR02R03CompPythia} shows that the ratio between the
measured jet spectra for radii of $R=0.2$ and $R=0.3$ is consistent with jet
production in vacuum for central and peripheral events. No sign of a modified
jet structure is observed between radii of $0.2$ and $0.3$ in the
ratio of the cross sections.
\Fig{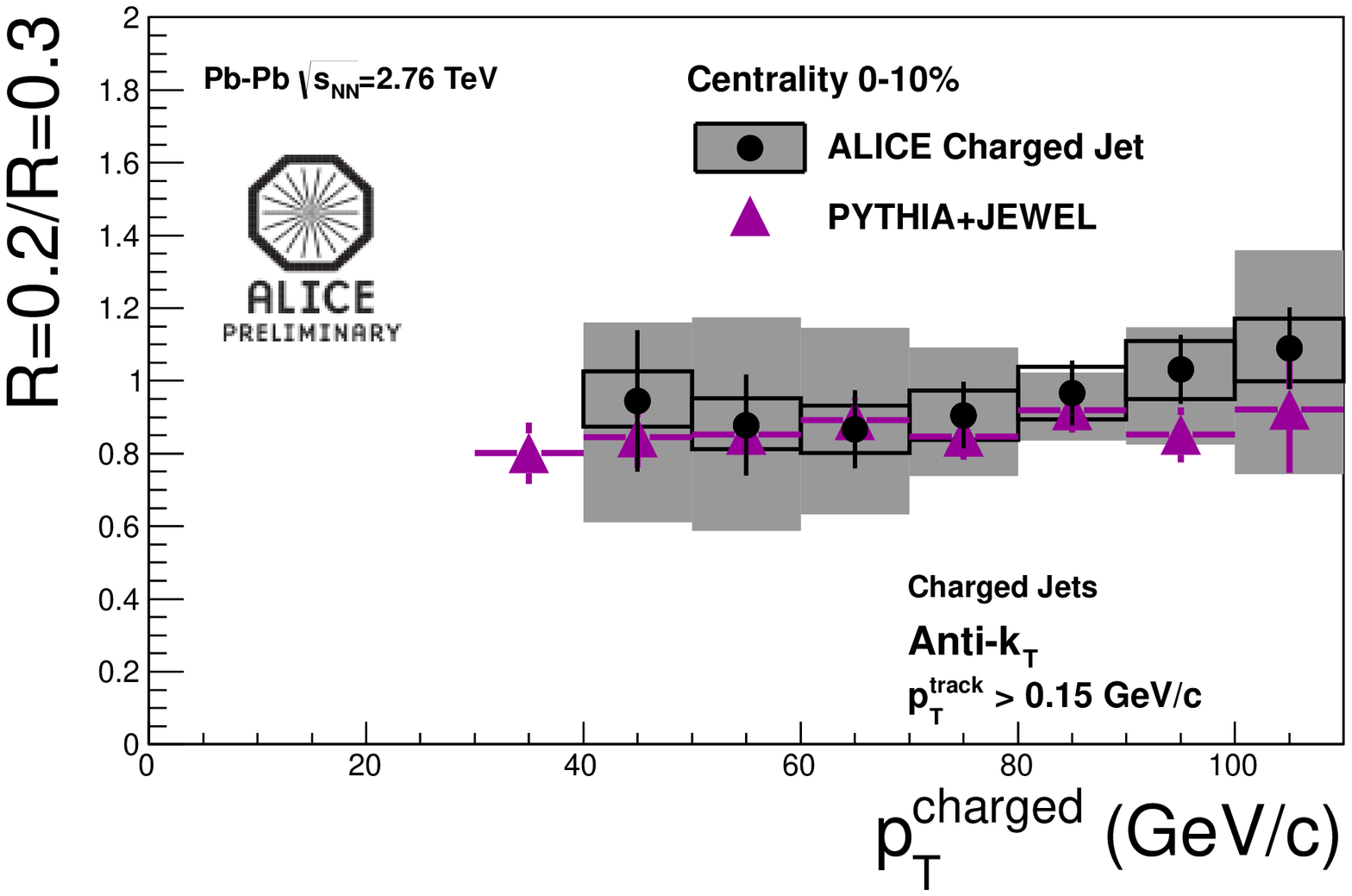}{fig:JEWEL}{width=0.68\linewidth}{
  Ratio between meastured cross-sections with radii $R=0.2$ and $R=0.3$ \PbPb
  compared to JEWEL energy loss MC.  }
The measured ratio of cross sections in \PbPb collisions is compared to the
JEWEL jet quenching MC \cite{Zapp:2011ek, zapp} in Figure \ref{fig:JEWEL}. A
good agreement is observed between the energy loss implementation of JEWEL and
the charged jet results from ALICE.


\bibliographystyle{elsarticle-num}
\bibliography{biblio}

\begin{thebibliography}{10}
\expandafter\ifx\csname url\endcsname\relax
  \def\url#1{\texttt{#1}}\fi
\expandafter\ifx\csname urlprefix\endcsname\relax\def\urlprefix{URL }\fi
\expandafter\ifx\csname href\endcsname\relax
  \def\href#1#2{#2} \def\path#1{#1}\fi

\bibitem{Cacciari2008b}
M.~Cacciari, G.~P. Salam, G.~Soyez, The anti-kt jet clustering algorithm, JHEP
  0804 (2008) 063.

\bibitem{Cacciari2006}
M.~Cacciari, G.~P. Salam, Dispelling the $n^3$ myth for the kt jet-finder,
  Phys.Lett.B 641 (2006) 57--61.
\newblock \href {http://arxiv.org/abs/hep-ph/0512210}
  {\path{arXiv:hep-ph/0512210}}.

\bibitem{Cacciari:2007fd}
M.~Cacciari, G.~P. Salam, {Pileup subtraction using jet areas}, Phys.Lett. B659
  (2008) 119--126.
\newblock \href {http://arxiv.org/abs/0707.1378} {\path{arXiv:0707.1378}},
  \href {http://dx.doi.org/10.1016/j.physletb.2007.09.077}
  {\path{doi:10.1016/j.physletb.2007.09.077}}.

\bibitem{Cacciari2010}
M.~Cacciari, J.~Rojo, G.~P. Salam, G.~Soyez, Jet reconstruction in heavy ion
  collisions, Eur.Phys.J.C 71 (2010) 1539.
\newblock \href {http://arxiv.org/abs/1010.1759} {\path{arXiv:1010.1759}}.

\bibitem{Abelev:2012ej}
B.~Abelev, et~al., {Measurement of Event Background Fluctuations for Charged
  Particle Jet Reconstruction in Pb-Pb collisions at $\sqrt{s_{NN}} = 2.76$
  TeV}, JHEP 1203 (2012) 053.
\newblock \href {http://arxiv.org/abs/1201.2423} {\path{arXiv:1201.2423}},
  \href {http://dx.doi.org/10.1007/JHEP03(2012)053}
  {\path{doi:10.1007/JHEP03(2012)053}}.

\bibitem{ALICE:2012ch}
M.~Verweij for~the collaboration, {Measurement of jet spectra in Pb-Pb
  collisions at $\sqrt{s_{NN}}$=2.76 TeV with the ALICE detector at the LHC
  }\href {http://arxiv.org/abs/1208.6169} {\path{arXiv:1208.6169}}.

\bibitem{RReedQM2012}
R.~J. Reed, Inclusive jet spectra in 2.76 TeV Pb-Pb collisions from the ALICE
  experiment, these proceedings.

\bibitem{Sjostrand2006}
T.~Sjostrand, S.~Mrenna, P.~Skands, {PYTHIA} 6.4 physics and manual, JHEP 05
  (2006) 026.
\newblock \href {http://arxiv.org/abs/hep-ph/0603175}
  {\path{arXiv:hep-ph/0603175}}.

\bibitem{Zapp:2011ek}
K.~C. Zapp, F.~Krauss, U.~A. Wiedemann, {Explaining jet quenching with
  perturbative QCD alone. }\href {http://arxiv.org/abs/1111.6838}
  {\path{arXiv:1111.6838}}.

\bibitem{zapp}
K.~C. Zapp, private communication.

\end{thebibliography}

%

\end{document}